\documentclass[12pt,english]{article}
\usepackage[margin=1in,bottom=1in]{geometry}

\usepackage[noblocks]{authblk} 

\usepackage{amsmath} 

\usepackage{upgreek} 

\usepackage{setspace}

\usepackage[document]{ragged2e} 
\usepackage{indentfirst} 

\usepackage[none]{hyphenat} 
\makeatletter
\let\X@old@caption\caption
\def\X@caption@minusone{\expandafter\advance\csname c@\@captype\endcsname-1 }
\def\X@caption@br[#1]#2{\X@old@caption[#1]{#2}\X@caption@minusone}
\def\X@caption@nobr#1{\X@old@caption{#1}\X@caption@minusone}
\def\caption{\@ifnextchar[\X@caption@br\X@caption@nobr}
\makeatother

\usepackage{booktabs} 
\usepackage[export]{adjustbox} 
\usepackage{float} 
\usepackage{color}

\usepackage{tablefootnote} 

\date{\vspace{-5ex}}

\usepackage{caption}
\usepackage{fancyhdr}

\begin{document}
\begin{spacing}{2.0}

\title{A Hidden Markov Movement Model for rapidly identifying behavioral states from animal tracks}
\author[1,6]{Kim Whoriskey} 
\author[1]{Marie Auger-M\'{e}th\'{e}}
\author[2]{Christoffer Moesgaard Albertsen}
\author[3]{Frederick G. Whoriskey}
\author[4]{Thomas R. Binder}
\author[5]{Charles C. Krueger}
\author[1]{Joanna Mills Flemming}
\affil[1]{Department of Mathematics and Statistics, Dalhousie University, Halifax, Nova Scotia B3H4R2 Canada}
\affil[2]{National Institute of Aquatic Resources, Technical University of Denmark, DK-2920 Charlottenlund, Denmark}
\affil[3]{Ocean Tracking Network, Dalhousie University, Halifax, Nova Scotia B3H 3Z1 Canada}
\affil[4]{Hammond Bay Biological Station, Department of Fisheries and Wildlife, Michigan State University, Millersburg, Michigan 49759 USA}
\affil[5]{Center for Systems Integration and Sustainability, Michigan State University, East Lansing, Michigan 48823-5243 USA}

\vspace{-20mm}
\maketitle

\thispagestyle{fancy} 
\renewcommand{\headrulewidth}{0pt} 
\fancyhf{}
\lhead{Hidden Markov Movement Model}
\lfoot{\textsuperscript{6} Email: kwhoriskey@dal.ca}
\cfoot{\thepage}

\begin{abstract}

\noindent 1. Electronic telemetry is frequently used to document animal movement through time. Methods that can identify underlying behaviors driving specific movement patterns can help us understand how and why animals use available space, thereby aiding conservation and management efforts. For aquatic animal tracking data with significant measurement error, a Bayesian state-space model called the first-Difference Correlated Random Walk with Switching (DCRWS) has often been used for this purpose. However, for aquatic animals, highly accurate tracking data of animal movement are now becoming more common.  

\noindent 2.  We developed a new Hidden Markov Model (HMM) for identifying behavioral states from animal tracks with negligible error, which we called the Hidden Markov Movement Model (HMMM). We implemented as the basis for the HMMM the process equation of the DCRWS, but we used the method of maximum likelihood and the R package \texttt{TMB} for rapid model fitting. 

\noindent 3.  We compared the HMMM to a modified version of the DCRWS for highly accurate tracks, the DCRWS$_{NOME}$, and to a common HMM for animal tracks fitted with the R package \texttt{moveHMM}. We show that the HMMM is both accurate and suitable for multiple species by fitting it to real tracks from a grey seal, lake trout, and blue shark, as well as to simulated data.

\noindent 4. The HMMM is a fast and reliable tool for making meaningful inference from animal movement data that is ideally suited for ecologists who want to use the popular DCRWS implementation for highly accurate tracking data. It additionally provides a groundwork for development of more complex modelling of animal movement with \texttt{TMB}. To facilitate its uptake, we make it available through the R package \texttt{swim}. 

\end{abstract}

\textbf{Key Words}: behavioral states; movement ecology, Ocean Tracking Network; Great Lakes Acoustic Telemetry Observation System; \texttt{swim}; \texttt{TMB}

\section*{Introduction}

Animals move to maximize their growth and to enhance their probability of survival and reproduction. Movement is therefore a critical animal behavior that reflects an animal's response to its current physical needs and environment (Hussey et al. 2015). Identifying these underlying drivers of animal movement (behavioral states) is required for understanding how and why animals use available space, and this knowledge informs the management and conservation of both species and ecosystems. In aquatic environments, where direct observation of animal movement and behavioral states is often impossible, researchers are rapidly expanding the use of electronic telemetry for documenting animal movement through time (Hussey et al. 2015). 

Satellite telemetry and acoustic positioning systems are the most common types of telemetry technology for estimating an aquatic animal's location in continuous space, and yield time series of locations along an animal's path, usually referred to as tracks. Inferring behavioral states from animal tracks is possible by assuming that different types of movement, and therefore behavioral states, can be reflected by changes in characteristics of an animal's path. For example, while foraging can often be characterized by a tortuous track, a more directed path may suggest travelling between habitats (e.g., Jonsen et al. 2005). 

Hidden Markov Models (HMMs) are a popular tool used to identify behavioral states from animal telemetry data with negligible error (e.g., Morales et al. 2004, Langrock et al. 2012). HMMs are a large class of models distinguished in the most general case by a set of observations that depend on a latent underlying Markov process (Zucchini and MacDonald, 2009). In the context of animal movement, the latent Markov process is used to model the discrete behavioral states of interest, while the set of observations follow a movement process that can also be Markovian. The observations can consist of either the location data (e.g., Jonsen et al. 2005) or metrics derived from the observed track, like turning angles and step lengths (e.g., Morales et al. 2004, Langrock et al. 2012). While current  HMMs can be fitted rapidly using maximum likelihood (ML) methods, with the exception of the formulation of Pedersen et al. (2011), they are unable to account for measurement error associated with the technology used to obtain animal tracks. 
For those tracks measured with error, state-space models (SSMs) provide a more accurate and reliable method for identifying behavioral states, but are typically fitted using comparatively slow Bayesian methods like Markov Chain Monte Carlo (MCMC) sampling because of large numbers of random effects (e.g., Jonsen et al. 2005, McClintock et al. 2012, Jonsen 2016). 
 Therefore, the ideal tool for identifying behavioral states from animal tracks should incorporate features of both HMM and SSM implementations, such that measurement error can be accounted for within a ML framework that keeps the computational burden of estimation relatively small. 

One particular SSM that has proven its utility through a wide range of applications on different species is the Bayesian first-Difference Correlated Random Walk with Switching (DCRWS) SSM of Jonsen et al. (2005). This model has been previously used to quantify foraging behavior in cetaceans (Bailey et al. 2009, Irvine et al. 2014), pinnipeds (Breed et al. 2009, Harwood et al. 2015),  turtles (Gonz\'{a}lez Carman et al. 2012, Hart et al. 2012), sea birds (Reid et al. 2014), and manta rays (Graham et al. 2012). Furthermore, it has been used to determine migration corridors (Prieto et al. 2014), estimate intraspecific competition (Breed et al. 2013), predator-prey relationships (Fitzpatrick et al. 2012), site fidelity (Block et al. 2011), and to inform management and conservation of protected regions (Block et al. 2011, Maxwell et al. 2011, Graham et al. 2012). 

Here we introduce a new HMM for estimating behavioral states from highly accurate animal tracks that is similar to the DCRWS, but does not account for measurement error. We directly implement the process equation of the DCRWS as the basis for our HMM, but we adjust the model for fitting within a ML framework to allow for rapid estimation. Model fitting and parameter estimation are performed using the R-package \texttt{TMB} (Kristensen et al. 2016), which has previously shown great promise for analyzing animal tracking data (Albertsen et al. 2015, Auger-M\'{e}th\'{e} et al. \textit{In press}). We make this model, which we entitle the Hidden Markov Movement Model (HMMM), available through the R package \texttt{swim} (see supplementary material). To demonstrate the accuracy and applicability of the HMMM, we apply it to simulated animal tracks and to real tracks from multiple aquatic species. We additionally compare our HMMM results to those obtained using its Bayesian counterpart and to results from the \texttt{moveHMM} package (Michelot et al. 2016). We assess the advantages and disadvantages of each approach by comparing their computational efficiency, accuracy, and sequences of behavioural states.

\section*{Methods}

\subsection*{The DCRWS Movement Process}

The DCRWS is a SSM that estimates the true locations, behavioral states, and parameters of a movement process from an Argos satellite system track  (Jonsen et al. 2005). Given the true location $\mathbf{x}_t$ at time $t$, the process equation of the DCRWS is a correlated random walk on the first differences of the true locations, $\mathbf{d}_t = \mathbf{x}_t - \mathbf{x}_{t-1}$:

\vspace{-7mm}
\begin{equation} \label{moveproc}
\mathbf{d}_t = \upgamma_{\mathrm{b}_{t-1}}\mathbf{T}(\uptheta_{\mathrm{b}_{t-1}})\mathbf{d}_{t-1} + \mathrm{N}_2(0, \Sigma)
\end{equation}
\begin{center}
$ \mathbf{T}(\uptheta_{\mathrm{b}_{t-1}}) = 
\begin{pmatrix}
cos(\uptheta_{\mathrm{b}_{t-1}}) & -sin(\uptheta_{\mathrm{b}_{t-1}})\\
sin(\uptheta_{\mathrm{b}_{t-1}}) & cos(\uptheta_{\mathrm{b}_{t-1}})
\end{pmatrix}
\hspace{10mm}
\Sigma = 
\begin{pmatrix}
\upsigma_{lon}^2 & \uprho\upsigma_{lon}\upsigma_{lat}\\
\uprho\upsigma_{lon}\upsigma_{lat} & \upsigma_{lat}^2
\end{pmatrix}
$
\end{center}

\noindent 
The stochastic term in the movement process is a bivariate Gaussian (N$_2$) with mean 0 and covariance matrix $\Sigma$, where $\upsigma_{lat}$ and $\upsigma_{lon}$ are the standard deviations in the latitude and longitude axes, respectively, and $\uprho$ is the correlation between the two axes. Like Breed et al. (2012), we assume here that $\uprho = 0$, implying that stochasticity in the latitude and longitude directions are independent of each other. The parameter $\mathbf{\upgamma}_{\mathrm{b}_{t-1}}$ describes the autocorrelation in both direction and speed, and $\mathbf{T}(\uptheta_{\mathrm{b}_{t-1}})$ is the rotational matrix through space given the turning angle $\mathbf{\uptheta}_{\mathrm{b}_{t-1}}$.
Multiple values are possible for $\mathbf{\upgamma}_{\mathrm{b}_{t-1}}$ and $\mathbf{\uptheta}_{\mathrm{b}_{t-1}}$, and these parameter values are dependent on the behavioral state at time $t-1$, i.e., $\mathrm{b}_{t-1}$. This dependence provides the mechanism for distinguishing between multiple behavioral states at each location. Typically of interest are two states: the first is directed movement characterized by travelling in the same direction ($\uptheta\approx0$) and at a similar, high speed ($\upgamma >0.5$), and the second is tortuous movement characterized by frequent course reversals ($\uptheta\approx\uppi$) at dissimilar, slower speeds ($\upgamma<0.5$). Parameter sets for each state are identified with the appropriate subscript, either 1 or 2.

\subsection*{The HMMM}

Our HMMM uses the movement process described by (\ref{moveproc}), but instead of utilizing a Bayesian framework like Jonsen et al. (2005), we employ a ML framework and fit the process equation as a HMM via \texttt{TMB}, which requires that the likelihood function be coded in the C++ programming language. The probability distribution of the movement process is conditional on the assumed behavioral states $\mathrm{b}_t$, and is given by

\vspace{-6mm}
\begin{equation} \label{switching}
f(\mathbf{d}_t | \mathrm{b}_{t-1}) \sim \textrm{N}_2(\upgamma_{\mathrm{b}_{t-1}}\mathbf{T}(\uptheta_{\mathrm{b}_{t-1}})\mathbf{d}_{t-1}, \Sigma) 
\end{equation}

The likelihood of the HMMM is that of a HMM (Zucchini and MacDonald 2009): 

\vspace{-6mm}
\begin{equation} \label{lik}
\updelta' \mathbf{P}(\mathbf{d}_1)\mathbf{A} \mathbf{P}(\mathbf{d}_2)\mathbf{A}  \cdot \cdot \cdot \mathbf{P}(\mathbf{d}_{t-1})\mathbf{A} \mathbf{P}(\mathbf{d}_t)\mathbf{1}
\end{equation}

\noindent Assuming two behavioral states, the $2\times1$ vector $\updelta$ contains the initial probabilities of being in each state. \textbf{A} is a $2 \times 2$ transition probability matrix containing the switching probabilities $\upalpha_{i,j}$ that describe the probability of switching from state $i$ at time $t-1$ to state $j$ at time $t$. Because the rows of \textbf{A} sum to 1, we need only estimate two switching probabilities instead of four; we choose to estimate $\alpha_{1,1}$ and $\alpha_{2,1}$. \textbf{P} is a $2 \times 2$ diagonal matrix with diagonal entries equal to ($f(\mathbf{d}_t | \mathrm{b}_{t-1} = 1)$, $f(\mathbf{d}_t | \mathrm{b}_{t-1} = 2)$), i.e. the probabilities of being at the observed locations given each behavioral state as described by the movement process. \textbf{1} is a $2 \times 1$ vector of ones. We estimate the parameters of the movement process directly from the likelihood within \texttt{TMB}, and then use the Viterbi algorithm to estimate the latent behavioral states (Zucchini and MacDonald 2009).

\subsection*{Data Analysis and Simulation Study}

To evaluate the performance of the HMMM, we compared it with two other approaches for estimating behavioral states from animal tracks with negligible measurement error. The first was the switching movement process described by (\ref{moveproc}) and (\ref{switching}), fitted using a Bayesian framework and Markov Chain Monte Carlo (MCMC) sampling via \texttt{rjags} (Plummer 2015). This first model is the DCRWS of Jonsen et al. (2005) modified such that it does not include a measurement equation, and therefore differs from the HMMM solely in implementation (i.e., Bayesian vs. ML inference). Hereafter we refer to it as the DCRWS$_{NOME}$. Although the DCRWS$_{NOME}$ has not been fitted before, implementations of the DCRWS for tracking data with minimal errors do exist (Jonsen 2016), and the DCRWS$_{NOME}$ is the most direct implementation of the DCRWS when no measurement error is assumed. 
To fit the DCRWS$_{NOME}$, we used a burn-in period of 40,000 samples, then sampled 20,000 from the posterior distribution but only kept every 20th sample (thinning). 
 We fitted and compared two MCMC chains to each track to check for convergence. All prior distributions were specified as in the R package \texttt{bsam} (Jonsen et al. 2015) that fits the original DCRWS model, with the exception of those for the error covariance matrix $\Sigma$. Instead, by setting $\uprho = 0$, which we believe is more appropriate, we were able to specify separate vague uniform priors on $\upsigma_{lon}$ and $\upsigma_{lat}$ as opposed to using the original Wishart prior on the entire matrix (Jonsen et al. 2005). Parameters and behavioral states were estimated as the posterior medians of the samples from the two chains combined.
We additionally fitted a HMM to the turning angles (rad) and step lengths (km) of the animal tracks with the R package \texttt{moveHMM} (Michelot et al. 2016), using a von Mises (mean $\upmu$ and concentration parameter $c$) and Weibull (shape $\uplambda$ and scale parameter $k$) distribution, respectively. Behavioral states were again identified via the Viterbi algorithm, using functions from \texttt{moveHMM}. 

We fitted these three models to three animal tracks: 1) a GPS track collected by a Sea Mammal Research Unit head-mounted Satellite Relay Data Logger (accurate GPS positions acquired when the head surfaces) deployed on an adult male grey seal (\textit{Halichoerus grypus}) at Kouchibouguac National Park, New Brunswick, in 2013; 2) an acoustic Vemco Positioning System (VPS; positions from triangulation of detections from multiple receivers in known locations; Smith 2013) track of an adult male lake trout (\textit{Salvelinus namaycush}) in northern Lake Huron in 2014; and 3) a light-based geolocation track recorded by an immature female blue shark (\textit{Prionace glauca}) tagged near Halifax, Nova Scotia, with a Wildlife Computers miniPAT tag in 2014. 
Since geolocation data can be error-prone, the track was processed with the Wildlife Computers (2015) GPE3 software (a SSM) to improve positioning accuracy by estimating the true blue shark locations, as suggested by the manufacturers. Although a SSM for geolocation data would have been the ideal approach for analyzing the blue shark data, our approach of fitting HMMs to true location SSM estimates has been previously adopted (e.g., Eckert et al. 2008).
 Because the data were collected in continuous-time but all three models assume underlying discrete-time Markov processes, we had to approximate the locations in discrete-time, and then assume that these were known. We linearly interpolated the data sets over time using a 6 hour, 15 minute, and 12 hour time step for the grey seal, lake trout, and blue shark data, respectively, yielding data sets with 1227, 2187, and 393 locations. Different time steps were required based on the different temporal resolutions of the tracks. 

Additionally, using the parameter estimates from the grey seal HMMM and DCRWS$_{NOME}$ fits (Table \ref{sealparms}), we conducted a simulation study to formally compare the accuracy of the HMMM and DCRWS$_{NOME}$, and compare their results with those obtained using \texttt{moveHMM}. We simulated 50 tracks corresponding to the HMMM from a known parameter set with turning angles $\uptheta_1 = 0, \uptheta_2 = \pi$, autocorrelation $\upgamma_1 = 0.8, \upgamma_2 = 0.05$, process error standard deviations $\upsigma_{lon} = 0.07, \upsigma_{lat} = 0.05$, and switching probabilities $\upalpha_{1,1} = 0.89, \upalpha_{2,2} = 0.80$. These two behavioral states are classically interpreted as transiting ($\uptheta_1, \upgamma_1$) and foraging ($\uptheta_2, \upgamma_2$). We then fitted the HMMM, \texttt{moveHMM}, and the DCRWS$_{NOME}$ to each simulated track and calculated the parameter estimates and interval measures of uncertainty for these estimates. For the HMMM and \texttt{moveHMM}, we used the 95\% confidence interval based on the standard error estimates. For the DCRWS$_{NOME}$, we determined the 95\% credible interval as the 2.5\% and 97.5\% quantiles of the posterior samples. We found the behavioral state error rate, i.e. the proportion of states that were incorrectly identified, for each model. We additionally calculated the root mean squared error (RMSE) for each parameter estimate $\hat{\Theta}$ from the HMMM and DCRWS$_{NOME}$ fits as 

\begin{equation} \label{rmse}
RMSE_{\Theta} = \bigg(\frac{1}{n}\displaystyle\sum\limits_{j=1}^n  (\hat{\Theta}-\Theta)^2\bigg)^{1/2}
\end{equation}

We were unable to calculate the RMSE for the \texttt{moveHMM} fits because the data were simulated according to the HMMM movement process and the \texttt{moveHMM} implementation does not involve the same parameters.

\section*{Results}

\subsection*{Identifying Behavioral States}

\begin{figure} [H]
\hspace*{-1cm}
\includegraphics[scale=0.65, center]{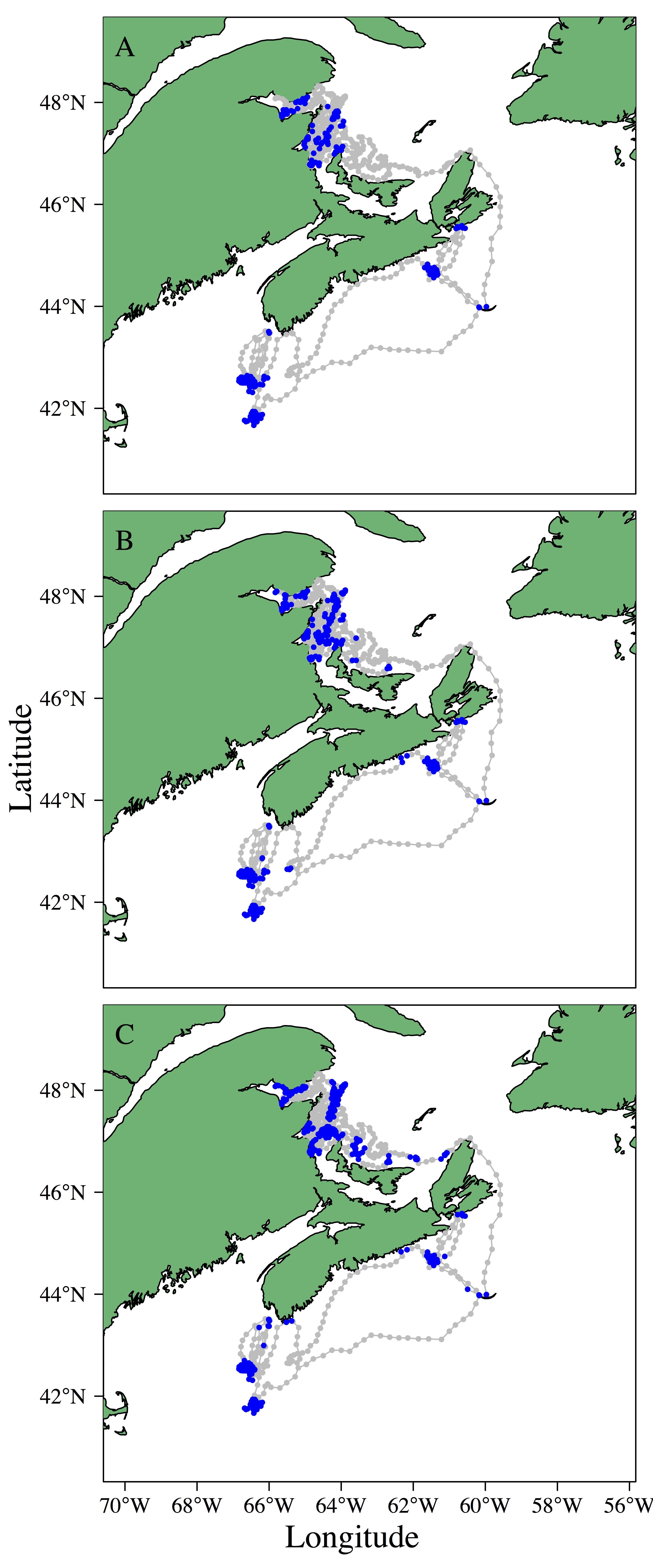} 
\caption{ Behavioral state results from the HMMM (panel A), DCRWS$_{NOME}$ (panel B), and \texttt{moveHMM} (panel C) models fitted to the grey seal track. Different behavioral states are indicated by grey (state 1) and blue (state 2) colors. \label{figureseal}}
\end{figure}

\begin{figure} [H]
\hspace*{-1cm}
\includegraphics[scale=0.65, center]{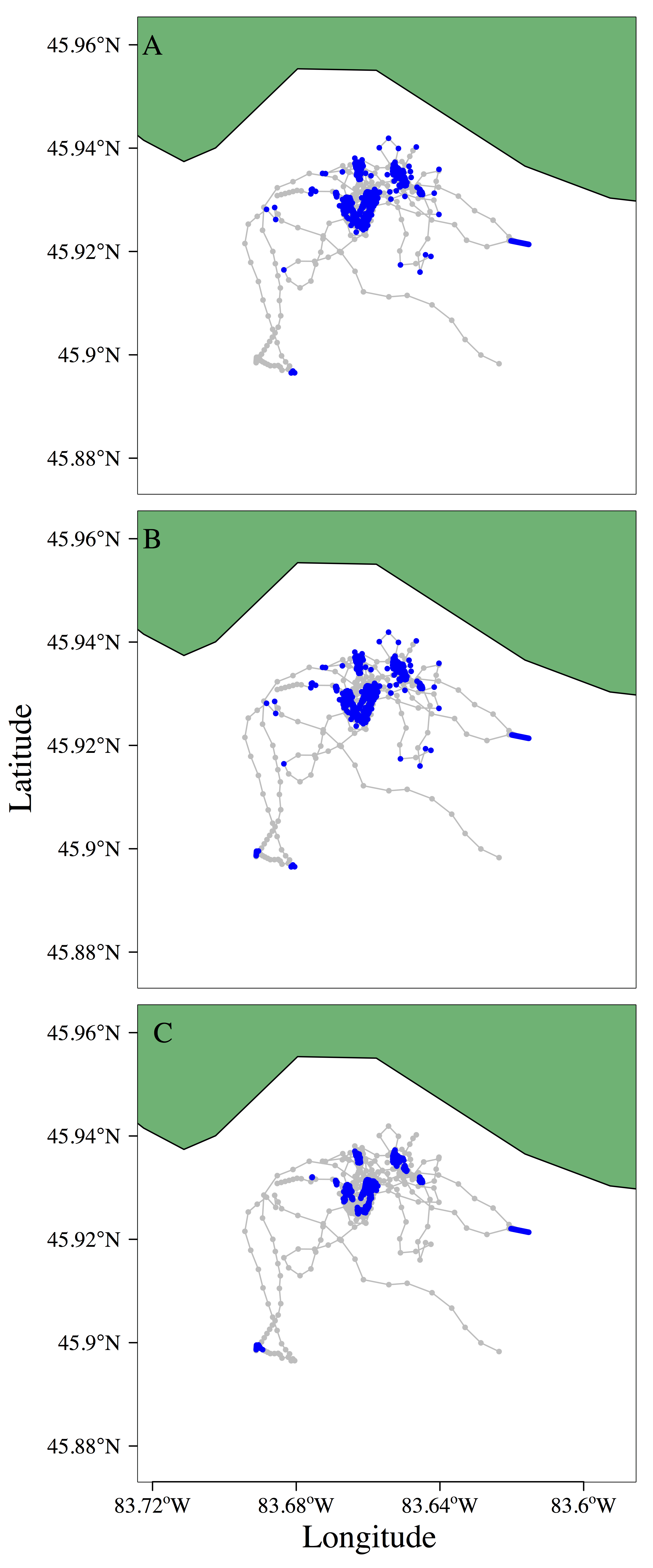} 
\caption{ Behavioral state results from the HMMM (panel A), DCRWS$_{NOME}$ (panel B), and \texttt{moveHMM} (panel C) models fitted to the lake trout track. Different behavioral states are indicated by grey (state 1) and blue (state 2) colors. \label{figurefish}}
\end{figure}

We applied the HMMM, the DCRWS$_{NOME}$, and \texttt{moveHMM} to a grey seal, lake trout, and blue shark track recorded by electronic telemetry with negligible measurement error. All three models performed similarly and identified two clearly distinct behavioral states for the grey seal and lake trout tracks (Figures \ref{figureseal},\ref{figurefish}). For both animals, HMMM and DCRWS$_{NOME}$ parameter estimates were similar except for the tortuous turning angle $\uptheta_2$, which was estimated at a similar distance from the number $\pi$ but in the opposite direction (i.e., while one was estimated turning slightly to the left, the other was estimated turning to the right; Tables \ref{sealparms}, \ref{fishparms}). This, along with the relatively large confidence intervals for $\uptheta_2$, is not unusual because together with small $\upgamma_2$, it suggests that the animal is exhibiting tortuous movement, in which case the mean turning angle does not have as much influence because the animal is more equally likely to travel in any direction. Switching probabilities ($\upalpha_{1,1}$ and $\upalpha_{2,1}$) were similar amongst all three models for the seal track. \texttt{moveHMM} estimated switching probabilities for the lake trout track different from the HMMM and DCRWS$_{NOME}$, although the estimated probabilities amongst all three models led to similar state estimates. For the seal track, the DCRWS$_{NOME}$ took 6.4 hours to fit, \texttt{moveHMM} took 0.9 seconds, and the HMMM took 0.06 seconds. For the lake trout data, the DCRWS$_{NOME}$ took 9.4 hours to fit, \texttt{moveHMM} took 1.8 seconds, and the HMMM took 0.16 seconds. 

\begin{figure} [H]
\hspace*{-1cm}
\includegraphics[scale=0.65, center]{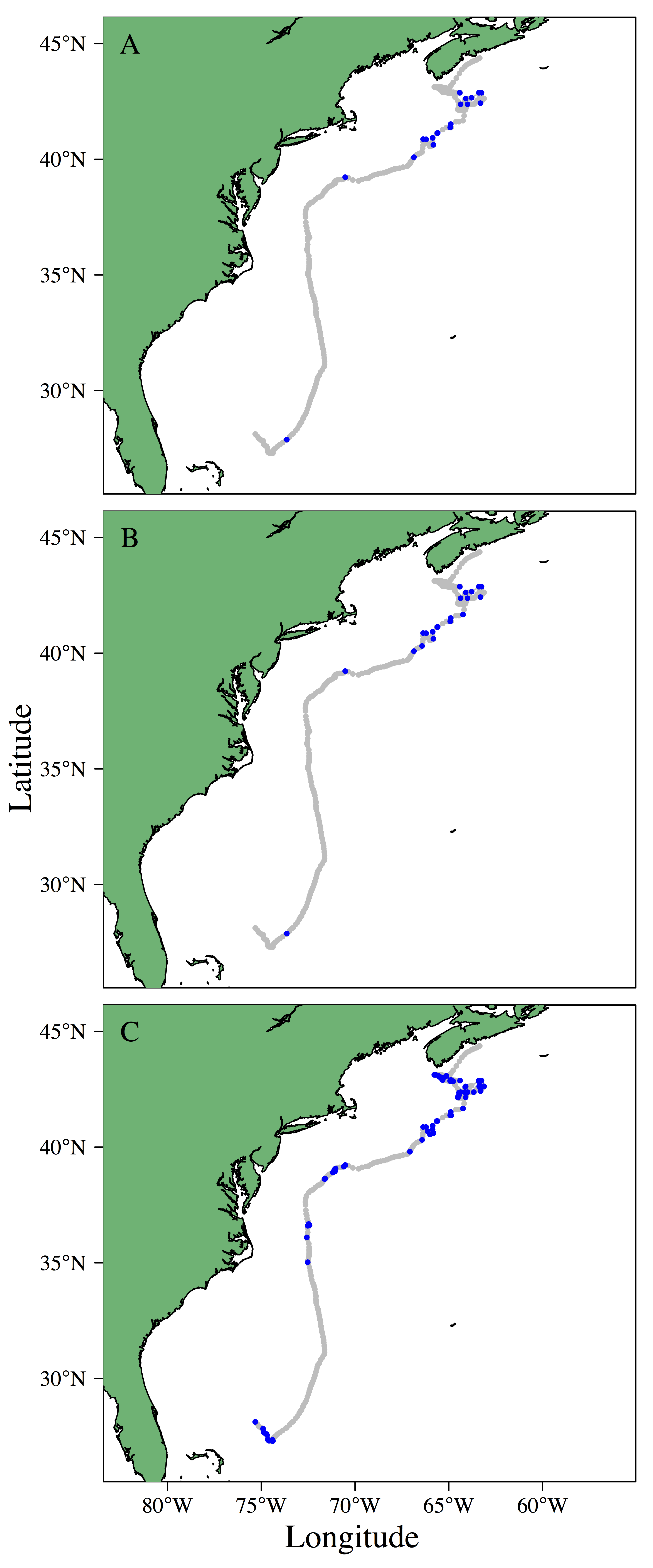} 
\caption{ Behavioral state results from the HMMM (panel A), DCRWS$_{NOME}$ (panel B), and \texttt{moveHMM} (panel C) models fitted to the blue shark track. Different behavioral states are indicated by grey (state 1) and blue (state 2) colors. \label{figureshark}}
\end{figure}

All three models identified two states from the blue shark track, although half of the switching probabilities estimated by \texttt{moveHMM} differed greatly from those estimated by the HMMM and DCRWS$_{NOME}$ (Table \ref{blueparms}), and this led to different decoded behavioral state sequences (Figure \ref{figureshark}). Specifically, all three models estimated a high probability of remaining in state 1, $\alpha_{1,1}$, but \texttt{moveHMM} estimated a low probability of switching from state 2 to state 1, $\alpha_{2,1}$ while the DCRWS$_{NOME}$ and HMMM estimated a high $\alpha_{2,1}$. The switching probabilities of the HMMM and DCRWS$_{NOME}$ therefore led to state sequences containing long stretches of state 1 interspersed with short (length 1 or 2) stretches of state 2. By contrast, \texttt{moveHMM} estimated a state sequence with longer stretches of both behavioral states. While the DCRWS$_{NOME}$ took 1.7 hours to fit to the blue shark track, \texttt{moveHMM} took 1.2 seconds, and the HMMM took 0.02 seconds. 

\subsection*{Simulation Study}

\begin{figure}[H] 
\hspace{-1cm}
\includegraphics{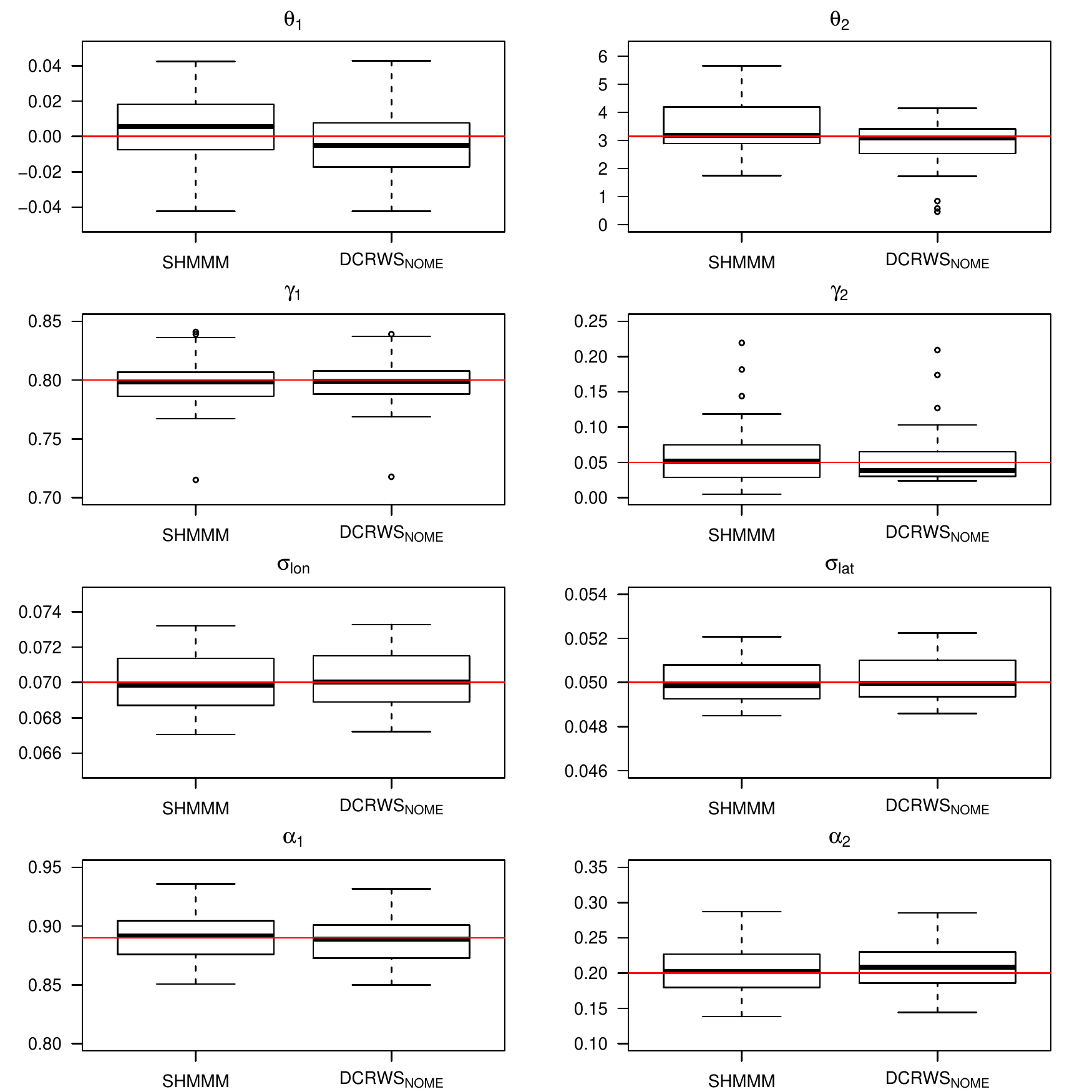}
\caption{ Boxplots of parameter estimates resulting from fitting the HMMM and the DCRWS$_{NOME}$ to 50 simulated tracks based on the grey seal application parameters. \label{figuresim}}
\end{figure}

We simulated 50 tracks from the HMMM with a specific parameter set (representative of the grey seal track) to test the accuracy of the HMMM compared to the DCRWS$_{NOME}$ and \texttt{moveHMM}. 
The HMMM and DCRWS$_{NOME}$ provided accurate estimates of the model parameters (Figure \ref{figuresim}), but the DCRWS$_{NOME}$ had a smaller average (over the parameters) RMSE (0.120 vs. 0.140; Table \ref{parmsim}). 
The RMSE for individual parameters were very similar (within 0.01) between the two models with the exception of $\uptheta_2$, where the RMSE of the DCRWS$_{NOME}$ was smaller by 0.149 (Table \ref{parmsim}). 
The DCRWS$_{NOME}$ additionally had the smallest behavioral state error rate (0.175) which differed from the HMMM and \texttt{moveHMM} by approximately 1.5\% (0.189) and 18.7\% (0.362), respectively. Finally, the average time needed to fit the DCRWS$_{NOME}$ was 5.10 hours, while \texttt{moveHMM} took 1.2 seconds and the HMMM took 0.08 seconds.

\section*{Discussion}

We have shown that the HMMM is a fast and reliable tool for estimating behavioral states from animal tracking data that contain negligible error. Our simulation study demonstrates the accuracy of the HMMM for estimating both the states and the model parameters. Our use of data from different species and derived by different electronic telemetry systems demonstrates the wide-ranging applicability of the HMMM. Coupled with the existing documentation of the DCRWS (more than 45 papers using this model), we suspect that the HMMM will be an easily interpretable tool for ecologists who are interested in implementing the DCRWS and have highly accurate data, which has an advantage over current methods with the DCRWS of being fast to fit (on the order of seconds) and avoiding convergence issues with MCMC samplers. HMMM implementation is available through the R package \texttt{swim}.

The HMMM, DCRWS$_{NOME}$, and \texttt{moveHMM} all identified two behavioral states from the grey seal track, consistent with previous analyses (Jonsen et al. 2005, Breed et al. 2009, Breed et al. 2011).
Grey seal tracks from Atlantic Canada typically exhibit clear bouts of both directed and tortuous movement, and have been previously analyzed with the DCRWS (e.g., Jonsen et al. 2005), therefore an Atlantic Canada grey seal GPS track provided an ideal test for our study. 
For grey seals, tortuous movement is classically interpreted as foraging, while directed movement is often regarded as travelling between foraging patches. The models identified several bouts of foraging behavior in the Northwest Atlantic and in the Gulf of Saint-Lawrence, specifically off the coasts of Nova Scotia, Prince Edward Island, New Brunswick, and Gasp\'{e}sie, areas of high biological productivity that are consistent with those previously identified as grey seal foraging areas (Breed et al. 2009).

With the HMMM, DCRWS$_{NOME}$, and \texttt{moveHMM}, we identified two behavioral states within the lake trout track. 
The Drummond Island lake trout population spawn primarily at nighttime on rock rubble reefs in association with submerged drumlins (Riley et al. 2014, Binder et al. 2015). Lake trout show multiple behaviors characterized by tortuous movement, including spawning on the reefs. For example, lake trout (particularly males) often aggregate on the spawning reefs in the weeks leading up to spawning, a behavior known as staging (Muir et al. 2012). Because egg surveys have verified that no spawning occurs in some locations where our models identified tortuous behavior (T. Binder, unpublished observations), we believe the models are distinguishing, more generally, reef and non-reef behaviors. Being able to mathematically distinguish between reef and non-reef behaviors can allow for identification of key lake trout habitats for conservation like spawning sites in places where direct observation is difficult. 
Furthermore, by building a dependence of the HMMM on one or more covariates, it may be possible to more acutely identify spawning behavior. For example, because the Drummond Island lake trout tend to spawn at night close to the substrate, time of day and lake trout depth (which is often recorded by positioning systems like the VPS) may provide sufficient additional information for the HMMM to distinguish spawning behavior from other reef-associated behaviors. One possible way to achieve this is by allowing the switching probabilities of the HMMM to depend on these covariates in a linear fashion (as in e.g., Bestley et al. 2012, Michelot et al. 2016). Additionally, an extension to the HMMM which could estimate more than two behavioural states may be able to distinguish reef from spawning behavior. We chose to model only two states so that we could more directly compare results of the HMMM to our implementation of the original DCRWS (the DCRWS$_{NOME}$); however, the HMMM should be directly extendible.

When fitted to the blue shark track, \texttt{moveHMM} produced different state sequences than the HMMM and DCRWS$_{NOME}$, as \texttt{moveHMM} estimated longer stretches of behavioral state 2 than either of the other models. This is likely because \texttt{moveHMM} models the distributions of the turning angles and step lengths calculated from an animal path, which is fundamentally different from the movement process of the HMMM and DCRWS$_{NOME}$. 
Furthermore, McClintock et al. (2014) showed that the continuous-time analog to the movement process introduced by Jonsen et al. (2005) and modeled by the HMMM has step lengths and bearings (turning angles) that are correlated, whereas the step lengths and bearings of the process modeled by McClintock et al. (2012) (close to that of \texttt{moveHMM}) are uncorrelated.
\texttt{moveHMM} identified two behaviors that were distinguished primarily by different step lengths, and therefore travelling speeds, with state 1 characterized by longer step lengths and faster speeds, and state 2 characterized by slower movement. The HMMM and DCRWS$_{NOME}$ identified two behaviors that are distinguished by high (state 1) and low (state 2) autocorrelations, or how related the speed at time $t$ is to the speed at time $t-1$. By modelling autocorrelation, the HMMM and DCRWS$_{NOME}$ are able to directly estimate persistence in animal movement, which reflects an animal's choice to move. It is possible that the shorter sequences of state 2 identified by the HMMM and DCRWS$_{NOME}$ resulted because the behaviours they are trying to estimate occur on a finer time scale than was modelled, which could make biological interpretation of these states difficult.

Our simulation study results suggested that while the DCRWS$_{NOME}$ was slightly more accurate than the HMMM, the difference was marginal. The two models performed very similarly while estimating model parameters with the exception of $\uptheta_2$, which the DCRWS$_{NOME}$ more accurately estimated. This result is likely explained by the rather informative priors on $\uptheta_1$ and $\upgamma_1$ when fitting the DCRWS$_{NOME}$.
 The DCRWS$_{NOME}$ also more accurately estimated the behavioral states, an unsurprising result because while the DCRWS$_{NOME}$  directly estimates these random effects from the posterior likelihood, the HMMM uses a post hoc global decoding algorithm (the Viterbi algorithm) to identify the most likely state at each location given the ML parameter estimates. Predictably, \texttt{moveHMM} had the highest behavioral state error rate, likely because it was fitted to simulated data from a movement process not equivalent to its own. Finally, the HMMM was the fastest model to fit, with \texttt{moveHMM} and DCRWS$_{NOME}$ taking on average 15 times and 229,500 times longer to fit than the HMMM, respectively.  Quicker fits of the DCRWS$_{NOME}$ may be achieved by reducing burn-in and sampling sizes of the MCMC, but they would still take orders of magnitude longer and may be less accurate. We chose these sizes based on prior experience with fitting the DCRWS, and to try to ensure convergence of the MCMC chains during the simulation study.

Our HMMM is a major advance in using \texttt{TMB} to solve animal movement problems. The HMMM is a HMM, and is therefore appropriate for location data with negligible error. Highly accurate data is becoming more common in the marine realm, and the HMMM, as implemented through the R package \texttt{swim}, provides a fast and reliable tool for making meaningful inference from animal movement data. Faster methods for analyzing data will become more important as larger data sets are collected. The HMMM therefore additionally provides a baseline method for movement modelling in \texttt{TMB} that can be further developed for more specific and nontrivial animal movement problems like determining relationships between movement and environmental covariates, or accounting for measurement error.

\section*{Acknowledgments} 
The grey seal and blue shark tracks were obtained under the approval of animal care protocols 12-64 and 13-002, respectively, from the University Committee on Laboratory Animals, of Dalhousie University's animal ethics committee. 
This research was funded by the Ocean Tracking Network (OTN), a Canadian Statistical Sciences Institute (CANSSI) Collaborative Team Project, a NSERC Discovery grant to Joanna Mills Flemming, and a Killam Predoctoral Scholarship to Kim Whoriskey. This research was also supported by the Great Lakes Fishery Commission by way of Great Lakes Restoration Initiative appropriations (GL-00E23010). We thank the OTN Bioprobe group, specifically Damian Lidgard, for access to the grey seal track, and Don Bowen and Shelley Lang for comments on the manuscript. We additionally thank Th\'{e}o Michelot for his advice on implementing \texttt{moveHMM}, and two anonymous reviewers for their time and constructive comments. This paper is contribution 25 of the Great Lakes Acoustic Telemetry Observation System (GLATOS). 
%
%
%
%

\section*{Literature Cited}

\noindent\hangindent=0.7cm Albertsen, C. M., K. Whoriskey, D. Yurkowski, A. Nielsen, and J. Mills Flemming. (2015). Fast fitting of non-Gaussian state-space models to animal movement data via Template Model Builder. \textit{Ecology} 96:2598-2604.

\noindent\hangindent=0.7cm Auger-M\'{e}th\'{e}, M., C. M. Albertsen, I. D. Jonsen, A. E. Derocher, D. C. Lidgard, K. R. Studholme, W. D. Bowen, G. T. Crossin, and J. Mills Flemming. (\textit{In press}). Spatiotemporal modelling of marine movement data using Template Model Builder. \textit{Marine Ecology Progress Series}. 


\noindent\hangindent=0.7cm Bailey, H., Mate, B. R., Palacios, D. M., Irvine, L., Bograd, S. J., Costa, D. P. (2009). Behavioural estimation of blue whale movements in the Northeast Pacific from state-space model analysis of satellite tracks. \textit{Endangered Species Research} 10:93-106

\noindent\hangindent=0.7cm Bestley, S. I. D. Jonsen, M. A. Hindell, C. Guinet, and J.-P. Charrassin. 2012. Integrative modelling of animal movement: incorporating in situ habitat and behavioural information for a migratory marine predator. \textit{Proceedings of the Royal Society B} 280: 20122262.

\noindent\hangindent=0.7cm Binder, T. R., H. T. Thompson, A. M. Muir, S. C. Riley, J. E. Marsden, C. R. Bronte, and C. C. Krueger. (2015). New insight into the spawning behavior of lake trout, \textit{Salvelinus namaycush}, from a recovering population in the Laurentian Great Lakes. \textit{Environmental Biology of Fishes} 98:173-181. 

\noindent\hangindent=0.7cm Block, B.  I. D. Jonsen, S. J. Jorgensen, A. J. Winship, S. A. Shaffer, S. J. Bograd, E. L. Hazen, D. G. Foley, G. A. Breed, A.-L. Harrison, J. E. Ganong, A. Swithenbank, M. Castleton, H. Dewar, B. R. Mate, G. L. Shillinger, K. M. Schaefer, S. R. Benson, M. J. Weise, R. W. Henry, and D. P. Costa. (2011). Tracking apex marine predator movements in a dynamic ocean. \textit{Nature} 475:86-90

\noindent\hangindent=0.7cm Breed, G. A., W. D. Bowen, and M. L. Leonard. (2011). Development of foraging strategies with age in a long-lived marine predator. \textit{Marine Ecology Progress Series} 431:267-279.

\noindent\hangindent=0.7cm Breed, G. A., W. D. Bowen, and M. L. Leonard. (2013). Behavioral signature of intraspecific competition and density dependence in colony-breeding marine predators. \textit{Ecology and Evolution} 3(11): 3838-3854

\noindent\hangindent=0.7cm Breed, G. A., D. P. Costa, I. D. Jonsen, P. W. Robinson, and J. Mills Flemming. (2012). State-space methods for more completely capturing behavioral dynamics from animal tracks. \textit{Ecological Modelling} 235-236:49-58 

\noindent\hangindent=0.7cm Breed, G. A., I. D. Jonsen, R. A. Myers, W. D. Bowen, and M. L. Leonard. (2009). Sex-specific, seasonal foraging tactics of adult grey seals (\textit{Halichoerus grypus}) revealed by state-space analysis. \textit{Ecology} 90:3209-3221. 

\noindent\hangindent=0.7cm Eckert, S. A. , J. E. Moore, D. C. Dunn, R. Sagarminaga van Buiten, K. L. Eckert, and P. N. Halpin. (2008). Modeling loggerhead turtle movement in the Mediterranean: Importance of body size and oceanography. \textit{Ecological Applications} 18:290-308. 


\noindent\hangindent=0.7cm Fitzpatrick, R., M. Thums, I. Bell, M. G. Meekan, J. D. Stevens, and A. Barnett. (2012). A Comparison of the Seasonal Movements of Tiger Sharks and Green Turtles Provides Insight into Their Predator-Prey Relationship. \textit{PLOS ONE} 7(12):e51927

\noindent\hangindent=0.7cm  Gonz\'{a}lez Carman, V., V. Falabella, S. Maxwell, D. Albareda, C. Campagna, and H. Mianzan. (2012). Revisiting the ontogenetic shift paradigm: The case of juvenile green turtles in the SW Atlantic. \textit{Journal of Experimental Marine Biology and Ecology} 429:64-72

\noindent\hangindent=0.7cm Graham, R. T., M. J. Witt, D. W. Castellanos, F. Remolina, S. Maxwell, B. J. Godley, and L. A. Hawkes. (2012). Satellite tracking of manta rays highlights challenges to their conservation. \textit{PLOS ONE} 7(5):e36834

\noindent\hangindent=0.7cm Hart K., M. M. Lamont, I. Fujisaki, A. D. Tucker, and R. R. Carthy. (2012). Common coastal foraging areas for loggerheads in the Gulf of Mexico: Opportunities for marine conservation. \textit{Biological Conservation} 145:185-194

\noindent\hangindent=0.7cm Harwood, L. A., T. G. Smith, J. C. Auld, H. Melling, and D. J. Yurkowski. (2015). Seasonal movements and diving of ringed seals, \textit{Pusa hispida}, in the Western Canadian Arctic, 1999-2001 and 2010-11. \textit{ARCTIC} 68(2):193-209 

\noindent\hangindent=0.7cm Hussey, N. E., S. T. Kessel, K. Aarestrup, S. J. Cooke, P. D. Cowley, A. T. Fisk, R. G. Harcourt, K. N. Holland, S. J. Iverson, J. F. Kocik, J. E. M. Flemming, and F. G. Whoriskey. (2015). Aquatic animal telemetry: A panoramic window into the underwater world. \textit{Science} 348:1255642.

\noindent\hangindent=0.7cm Irvine, L. M., B. R. Mate, M. H. Winsor, D. M. Palacios, S. J. Bograd, D. P. Costa, and H. Bailey. (2014). Spatial and Temporal Occurrence of Blue Whales off the U.S. West Coast, with Implications for Management. \textit{PLOS ONE} 9:e102959

\noindent\hangindent=0.7cm Jonsen, I. (2016). Joint estimation over multiple individuals improves behavioral state inference from animal movement data. \textit{Scientific Reports} 6:20625. 

\noindent\hangindent=0.7cm Jonsen, I. D., J. Mills Flemming, and R. A. Myers. (2005). Robust state-space modeling of animal movement data. \textit{Ecology} 86:2874-2880.

\noindent\hangindent=0.7cm Jonsen, I. (with contributions from) S. Luque, A. Winship and M.W. Pedersen (2015). bsam: Bayesian state-space models for animal movement. R package version 0.43.1. http://www.r-project.org

\noindent\hangindent=0.7cm Kristensen, K., A. Nielsen, C. W. Berg, and H. Skaug. (2016). TMB: Automatic Differentiation and Laplace Approximation. \textit{Journal of Statistical Software} 70 (5):1-21. doi:10.18637/jss.v070.i05

\noindent\hangindent=0.7cm Langrock, R., R. King, J. Matthiopoulos, L. Thomas, D. Fortin, and J. M. Morales. (2012). Flexible and practical modeling of animal telemetry data: hidden Markov models and extensions. \textit{Ecology} 93:2336-2342.

\noindent\hangindent=0.7cm Maxwell, S. M., G. A. Breed, B. A. Nickel, J. Makanga-Bahouna, E. Pemo-Makaya, R. J. Parnell, A. Formia, S. Ngouessono, B. J. Godley, D. P. Costa, M. J. Witt, and M. S. Coyne. (2011). Using satellite tracking to optimize protection of long-lived marine species: Olive ridley sea turtle conservation in Central Africa. \textit{PLOS ONE} 6(5):e19905

\noindent\hangindent=0.7cm McClintock, B. T., D. S. Johnson, M. B. Hooten, J. M. Ver Hoef, and J. M. Morales. 2014. When to be discrete: the importance of time formulation in understanding animal movement. \textit{Movement Ecology} 2:21.

\noindent\hangindent=0.7cm McClintock, B. T., R. King, L. Thomas, J. Matthiopoulos, B. J. McConnell, and J. M. Morales. (2012). A general discrete-time modeling framework for animal movement using multistate random walks. \textit{Ecological Monographs} 82:335-349.

\noindent\hangindent=0.7cm Michelot, T., R. Langrock, and T. A. Patterson. (2016). moveHMM: An R package for the statistical modelling of animal movement data using hidden Markov models. \textit{Methods in Ecology and Evolution}. doi:10.1111/2041-210X.12578

\noindent\hangindent=0.7cm Morales, J. M., D. T. Haydon, J. Frair, K. E. Holsinger, and J. M. Fryxell. (2004). Extracting more out of relocation data: Building movement models as mixtures of random walks. \textit{Ecology} 85:2436-2445. 

\noindent\hangindent=0.7cm Muir, A. M., C. T. Blackie, J. E. Marsden, and C. C. Krueger. (2012). Lake charr \textit{Salvelinus namaycush} spawning behaviour: New field observations and a review of current knowledge. \textit{Reviews in Fish Biology and Fisheries} 22:575-593. 


\noindent\hangindent=0.7cm Pedersen, M. W., T. A. Patterson, U. H. Thygesen and H. Madsen M. (2011). Estimating animal behavior and residency from movement data. \textit{Oikos} 120:1281-1290.

\noindent\hangindent=0.7cm Plummer, M. 2015. rjags: Bayesian Graphical Models using MCMC. R package version 3-15. http://CRAN.R-project.org/package=rjags

\noindent\hangindent=0.7cm Prieto, R., M. A. Silva, G. T. Waring, and J. M. A. Gon\c{c}alves. (2014). Sei whale movements and behaviour in the North Atlantic inferred from satellite telemetry. \textit{Endangered Species Research} 26:103-113. 

\noindent\hangindent=0.7cm Reid, T. A., R. A. Ronconi, R. J. Cuthbert, and P. G. Ryan. (2014). The summer foraging ranges of adult spectacled petrels \textit{Procellaria conspicillata}. \textit{Antarctic Science} 26(1):23-32. 

\noindent\hangindent=0.7cm Riley, S. C., T. R. Binder, N. J. Wattrus, M. D. Faust, J. Janssen, J. Menzies, J. E. Marsden, M. P. Ebener, C. R. Bronte, J. X. He, T. R. Tucker, M. J. Hansen, H. T. Thompson, A. M. Muir, and C. C. Krueger. (2014). Lake trout in northern Lake Huron spawn on submerged drumlins. \textit{Journal of Great Lakes Research} 40:415-420.

\noindent\hangindent=0.7cm Smith, F. (2013). Understanding HPE in the VEMCO Positioning System (VPS). VEMCO Document \#: DOC-005457-01. Available from http://vemco.com/

\noindent\hangindent=0.7cm Wildlife Computers. (2015). Data portal's location processing (GPE3 \& FastLoc-GPS) user guide. Available from http://wildlifecomputers.com/

\noindent\hangindent=0.7cm Zucchini, W., and I. MacDonald. (2009). Hidden Markov Models for Time Series: An Introduction Using R. Boca Raton, FL: Chapman \& Hall/CRC.

\newpage
\section*{Tables}

\begin{table}[H]
\caption{Parameter estimates from three models fitted to a grey seal track. Lower and upper columns are the lower and upper bound of 95\% uncertainty intervals around the estimates. These correspond to 95\% confidence intervals for the HMMM and \texttt{moveHMM}, and 95\% credible intervals for the DCRWS$_{NOME}$. The only two parameters in common between all three models are the switching probabilities, $\upalpha_{1,1}$ and $\upalpha_{2,1}$.  \label{sealparms}}
\centering
\hspace*{-1.7cm}
\begin{tabular}{ccccccccccc}
\toprule
Parameter & \multicolumn{3}{c}{HMMM} & \multicolumn{3}{c}{DCRWS$_{NOME}$} & Parameter & \multicolumn{3}{c}{moveHMM}\\
&  Estimate & Lower & Upper & Estimate & Lower & Upper & & Estimate & Lower & Upper\\
\hline
$\uptheta_1$ &           0.022   &   -0.023  &  0.066 &   -0.017 &  -0.060  &  0.027  &   $\upmu_1$       &  -0.010   &   -0.058  &    0.038\\
$\uptheta_2$ &           4.662   &   2.441    &  5.835 &   1.831  &  0.275   &  5.980  &   $\upmu_2$       &   0.495   &   -0.358  &   1.348\\
$\upgamma_1$ &       0.805   &   0.753    &  0.848 &   0.805  &  0.759   &  0.849  &   $c_1$               &   0.685    &   0.640   &   0.730\\
$\upgamma_2$ &       0.055   &  0.013     &  0.201 &   0.048  &  0.003   & 0.128   &   $c_2$               &   0.069    &    0.002  &   0.135\\
$\upsigma_{lon}$ &    0.071   &   0.068    &   0.074 &   0.071  &  0.068   & 0.074  &   $\uplambda_1$ &   2.185    &  1.977   &   2.393\\
$\upsigma_{lat}$ &    0.050    &   0.048    &   0.053 &   0.050  &  0.048   & 0.053  &   $\uplambda_2$ &   0.816    &  0.757    &  0.875\\
		        &                  &                &             &              &              &             &  $k_1$                  &  15.342   & 14.381   & 16.304\\
		        &                  &                &             &              &              &             &  $k_2$                  &  3.487    &   2.878    &  4.097\\
$\upalpha_{1,1}$ &   0.890     &   0.827    &   0.932  &  0.885 &  0.835   &  0.929 &                             &  0.876    &   0.842     & 0.910\\
$\upalpha_{2,1}$ &   0.198     &   0.133    &   0.285  &  0.204  &  0.141  &  0.292  &                            &  0.111     &   0.090     &  0.158\\
\bottomrule
\end{tabular}
\hspace*{-1.7cm}
\end{table}

\begin{table}[H]
\caption{Parameter estimates from three models fitted to a lake trout track. Lower and upper columns are the lower and upper bound of 95\% uncertainty intervals around the estimates. These correspond to 95\% confidence intervals for the HMMM and \texttt{moveHMM}, and 95\% credible intervals for the DCRWS$_{NOME}$. The only two parameters in common between all three models are the switching probabilities, $\upalpha_{1,1}$ and $\upalpha_{2,1}$.  \label{fishparms}}
\centering
\hspace*{-1.7cm}
\begin{tabular}{ccccccccccc}
\toprule
Parameter & \multicolumn{3}{c}{HMMM} & \multicolumn{3}{c}{DCRWS$_{NOME}$} & Parameter & \multicolumn{3}{c}{moveHMM}\\
&  Estimate & Lower & Upper & Estimate & Lower & Upper & & Estimate & Lower & Upper\\
\hline
$\uptheta_1$ &          -0.118     &    -0.155  &  -0.082  &  0.119     &  0.084   &  0.155    &  $\upmu_1$       & 0.021   & -0.041  &  0.083\\
$\uptheta_2$ &             2.687   &     2.277  &  3.113    &   3.603   &  3.206   &   4.088   &  $\upmu_2$       &  -0.746 & -2.042  &  0.447\\
$\upgamma_1$ &        0.821    &    0.786   &   0.851  &   0.821    &  0.788   &   0.853   &  $c_1$               &  3.123  &  2.488  &  3.763\\
$\upgamma_2$ &         0.128   &   0.083    &   0.191  &   0.123    &  0.075   &   0.177   &  $c_2$               &  0.113  &  0.033  &  0.238\\
$\upsigma_{lon}$ &      0.001   &   0.001    &   0.001  &   0.001    &   0.001  &   0.001   &  $\uplambda_1$ & 2.324  &  2.119  &  2.550\\
$\upsigma_{lat}$ &       0.001   &   0.001    &   0.001  &   0.001    &   0.001  &   0.001   &  $\uplambda_2$ &  0.838 &  0.786  &  0.894\\
		        &                  &                &             &              &              &             &        $k_1$                   & 16.128 &15.274 & 17.029\\
		        &                  &                &             &              &              &             &         $k_2$                  & 4.084  &  3.621  &  4.606\\
$\upalpha_{1,1}$ &    0.645     &   0.578    &  0.707   &  0.643   &  0.576      &   0.705  &                            & 0.853  &  0.811   &  0.887\\
$\upalpha_{2,1}$ &    0.288     &   0.212    &  0.377   & 0.289    &  0.214      &   0.384  &                            & 0.102  &  0.077   & 0.132 \\
\bottomrule
\end{tabular}
\hspace*{-1.7cm}

\end{table}

\begin{table}[H]
\caption{Parameter estimates from three models fitted to a blue shark track. Lower and upper columns are the lower and upper bound of 95\% uncertainty intervals around the estimates. These correspond to 95\% confidence behaviorintervals for the HMMM and \texttt{moveHMM}, and 95\% credible intervals for the DCRWS$_{NOME}$. The only two parameters in common between all three models are the switching probabilities, $\upalpha_{1,1}$ and $\upalpha_{2,1}$. Because this track had some step lengths equal to zero, the two parameters $\zeta_1$ and $\zeta_2$ were used to estimate zero-inflation for each behavior when using \texttt{moveHMM}. \label{blueparms}}
\centering
\hspace*{-1.7cm}
\begin{tabular}{ccccccccccc}
\toprule
Parameter & \multicolumn{3}{c}{HMMM} & \multicolumn{3}{c}{DCRWS$_{NOME}$} & Parameter & \multicolumn{3}{c}{moveHMM}\\
&  Estimate & Lower & Upper & Estimate & Lower & Upper & & Estimate & Lower & Upper\\
\hline
$\uptheta_1$ &          -0.021    &  -0.070     &  0.027   &  0.013   & -0.040  &   0.062    &  $\upmu_1$       &  -0.003  & -0.025   &  0.019 \\
$\uptheta_2$ &            0.528    &   0.232    &  1.131  &   -0.881  & -0.006  &   0.157    &  $\upmu_2$       &  0.013   & -0.273   &  0.300  \\
$\upgamma_1$ &        0.923    &   0.846    &  0.963  &   0.932   &  0.873   &  0.987   &  $c_1$               &   40.323  &  30.556 & 50.107\\
$\upgamma_2$ &         0.289   &   0.199    &  0.400  &   0.303   &  0.188   &  0.423   &  $c_2$               &    0.949   &  0.616   & 1.302  \\
$\upsigma_{lon}$ &      0.045   &   0.042    &  0.049 &    0.046   &  0.043    &  0.049   &  $\uplambda_1$ &  1.806   &  1.608   &   2.029 \\ 
$\upsigma_{lat}$ &       0.042   &   0.039    &  0.045  &   0.042   &  0.038    &  0.045   &  $\uplambda_2$ &  1.069   &  0.927   &  1.232 \\
		        &                  &                &             &              &              &             &        $k_1$                 &  11.816 &  10.872  & 12.842  \\
		        &                  &                &             &              &              &             &         $k_2$                &  6.610   &   5.255   &  8.314  \\
		        &                  &                &             &              &              &             &         $\zeta_1$          &   0.029  &   0.014   &  0.059\\
		        &                  &                &             &              &              &             &         $\zeta_2$          &  0.035   &   0.013   &  0.091\\
$\upalpha_{1,1}$ &    0.904    &   0.794     &   0.958  &  0.880    & 0.732    &  0.955  &                             &  0.841   &    0.778  &   0.888  \\
$\upalpha_{2,1}$ &    0.722    &   0.385     &   0.915  &  0.742    &  0.437   &  0.925  &                             &  0.320   &   0.211   &   0.454  \\
\bottomrule
\end{tabular}
\hspace*{-1.7cm}
\end{table}

\begin{table} [H]
\caption{Parameter results from the simulation study (n=50) based on grey seal application parameters comparing the HMMM to the DCRWS$_{NOME}$. The Lower and Upper columns correspond to the 95\% confidence and credible intervals for the HMMM and the DCRWS$_{NOME}$, respectively. The Estimate, Lower, and Upper columns are averages over all simulations. The RMSE columns contain the root mean squared errors. \label{parmsim}}
\centering
\hspace*{-1cm}
\begin{tabular}{cccccccccc}
\toprule
Parameter & True Value & \multicolumn{4}{c}{HMMM} & \multicolumn{4}{c}{DCRWS$_{NOME}$}\\
                 &                    &  Estimate & Lower & Upper & RMSE & Estimate & Lower & Upper & RMSE\\
\hline
$\uptheta_1$ &     0            & 0.004   &  -0.035  &  0.042 &  0.020  &   -0.004   &  -0.043 &   0.035   & 0.020 \\
$\uptheta_2$ &  $\pi$         & 3.433   &  1.599   &   4.730 &  0.978  &  2.931     &   0.483  &   5.400  & 0.829\\
$\upgamma_1$ &  0.80      & 0.797   &  0.752   &   0.836 &  0.021  &  0.797      &   0.756   &  0.839  & 0.021\\
$\upgamma_2$ &   0.05     & 0.060     & 0.018  &  0.321 &  0.044  &   0.052     &   0.007   &  0.138 & 0.037\\
$\upsigma_{lon}$ & 0.07    & 0.070    &  0.067   &  0.073 &  0.002  &   0.070    &  0.067    & 0.073   & 0.002\\
$\upsigma_{lat}$ & 0.05     & 0.050    &  0.048   &  0.052 &  0.001  &  0.050     &0.048     &0.052     & 0.001\\
$\upalpha_{1,1}$ & 0.89     &  0.890   &  0.844  &   0.924 &  0.018  &   0.886      &0.842     & 0.922  & 0.018\\
$\upalpha_{2,1}$ & 0.20     & 0.205   &  0.140   & 0.290   &  0.034  &   0.208      &0.143     &0.293   & 0.033\\
\bottomrule
\end{tabular}
\hspace*{-1cm}
\end{table}

\newpage
\section*{Figures}

\end{spacing}
\end{document}